%% file: iclr2025_conference.tex
\documentclass{article}
\usepackage{iclr2025_conference,times}
\input{math_commands.tex}

\usepackage{algorithm}
\usepackage{algpseudocode}
\usepackage{amsmath}
\usepackage{booktabs}
\usepackage{graphicx}
\usepackage{natbib}
\usepackage{hyperref}

\title{IVE: Enhanced Probabilistic Forecasting of Intraday Volume Ratio with Transformers}

\author{
  Hanwool Lee \\
  Shinhan Securities, South Korea \\
  \texttt{gksdnf424@gmail.com} 
  \And
  Heehwan Park\thanks{This research was performed while the author was at NCSOFT} \\
  NCSOFT, South Korea \\
  \texttt{heehwan.park@gmail.com}
}

\iclrfinalcopy % Camera-ready 버전일 경우 주석 해제

\begin{document}

\maketitle

\begin{abstract}
This paper presents a new approach to volume ratio prediction in financial markets, specifically targeting the execution of Volume-Weighted Average Price (VWAP) strategies. Recognizing the importance of accurate volume profile forecasting, our research leverages the Transformer architecture to predict intraday volume ratio at a one-minute scale. We diverge from prior models that use log-transformed volume or turnover rates, instead opting for a prediction model that accounts for the intraday volume ratio's high variability, stabilized via log-normal transformation. Our input data incorporates not only the statistical properties of volume but also external volume-related features, absolute time information, and stock-specific characteristics to enhance prediction accuracy. The model structure includes an encoder-decoder Transformer architecture with a distribution head for greedy sampling, optimizing performance on high-liquidity stocks across both Korean and American markets. We extend the capabilities of our model beyond point prediction by introducing probabilistic forecasting that captures the mean and standard deviation of volume ratios, enabling the anticipation of significant intraday volume spikes. Furthermore, an agent with a simple trading logic demonstrates the practical application of our model through live trading tests in the Korean market, outperforming VWAP benchmarks over a period of two and a half months. Our findings underscore the potential of Transformer-based probabilistic models for volume ratio prediction and pave the way for future research advancements in this domain.
\end{abstract}

\section{Introduction}
In the dynamic realm of financial markets, precise execution strategies, notably the Volume-Weighted Average Price (VWAP), are essential~\cite{almgren2001optimal, Kato_2015}. Success in VWAP hinges on accurate volume ratio predictions, a complex challenge given market unpredictability. Addressing this, our research presents a new approach, leveraging the Transformer architecture~\cite{vaswani2023attentionneed} for high accuracy in intraday volume ratio predictions at a minute-level granularity.

Diverging from traditional models~\cite{ma2021predictingdailytradingvolume, Liu2016IntradayVP} reliant on log-transformed volumes or turnover rates, our model incorporates a log-normal transformation to counter the high variability of intraday volume ratios. We enrich our model with a diverse dataset, including external volume-related features, temporal information, and stock-specific characteristics, enhancing predictive accuracy.

Our model's architecture features an encoder-decoder Transformer structure with a distribution head for optimal performance on high-liquidity stocks in various markets, including Korean and American markets. An innovation is the introduction of probabilistic forecasting~\cite{lakshminarayanan2017simplescalablepredictiveuncertainty}, allowing for the estimation of volume ratio means and standard deviations. This capability is crucial for navigating market volatility and executing effective VWAP strategies.

The practical applicability of our model is demonstrated through live trading tests in the Korean market, where our approach outperformed VWAP benchmarks, validating its effectiveness and potential in real-world trading scenarios.

In summary, this paper makes three principal contributions to volume ratio prediction in financial markets:
\begin{enumerate}
    \item \textbf{Transformer-Based Volume Ratio Prediction Model}: This model achieves new levels of accuracy in volume ratio prediction, utilizing advanced machine learning techniques.
    \item \textbf{Probabilistic Forecasting Implementation}: Our research introduces probabilistic forecasting in volume ratio prediction, enabling improved anticipation of market movements.
    \item \textbf{Practical Trading Application}: The model's effectiveness is proven through live trading tests, outperforming established VWAP benchmarks.
\end{enumerate}

These contributions mark an advancement in financial modeling, integrating sophisticated machine learning techniques to enhance trading strategies.

\section{Backgrounds}

\subsection{Optimal Execution and VWAP}
Optimal execution is a pivotal strategy in financial trading, aimed at securing the best possible outcome in trading securities like stocks, primarily to minimize transaction costs. This method has traditionally been implemented through a trader's intuition or by mathematical modeling of the Limit Order Book dynamics, utilizing principles like Dynamic Programming and the Hamilton–Jacobi–Bellman equation~\cite{donnelly2022optimal}.

A key performance measure in optimal execution strategies is the Volume Weighted Average Price (VWAP). VWAP, influenced by the actions of all market participants, measures the average transaction price of all trades over a time period. The challenge in targeting VWAP stems from its dependency on the uncertain actions of other market participants and the inherent randomness in future market order-flows~\cite{Kato_2015}.

An emerging focus in optimal execution, especially within VWAP strategies, is the importance of precise, short-term predictions. The ability to accurately forecast market conditions on smaller time scales, such as minute-by-minute changes, is crucial for effective implementation of these strategies. This precision allows for more efficient order execution, better aligning with the constantly fluctuating market conditions~\cite{dang2013dynamic}.

Various VWAP strategies have been developed to follow this benchmark, emphasizing mathematical modeling of factors like the percentage of Volume (POV), percentage of cumulative volume (POCV), and turnover rate. These models aim to distribute the target quantity appropriately at each time point, considering the short-term market dynamics~\cite{Kato_2015}.

In VWAP-conforming optimal execution strategies, breaking down large orders into smaller ones for execution over shorter periods at better than average market prices is a key tactic. This method is particularly beneficial for large-scale orders, as it reduces market impact and uses price volatility to minimize transaction costs~\cite{Kato_2015}.

\subsection{Intraday Volume Prediction}
The U-shaped intraday volume pattern has long been observed, and statistical modeling has historically been used to predict this U-shape and the volume curve~\cite{jonathan1986theory}. Past studies have used statistical modeling techniques (e.g., ARIMA) and traditional machine learning approaches for prediction at different scales, including 50-minute, 10-minute, and up to 5-minute intervals~\cite{Graczyk_2016}.

These studies have indicated that gradient-based methodologies can be effective for precise predictions, and RNN-based approaches have been successfully implemented on a 50-minute scale. However, RNNs have limitations due to gradient vanishing, making them less suitable for shorter scales~\cite{Libman2019}.

In this paper, we effectively perform minute-scale probabilistic intraday volume forecasting using a Transformer-based structure. This approach allows for the construction of simple volume distribution strategies and superior performance in VWAP target strategies~\cite{Graczyk_2016}.

\section{Methodology and Experiments}

\subsection{Data Creation}
In advancing the methodology for trading volume prediction, this study diverges from traditional approaches that predominantly leveraged normalized volume or turnover rates. Recognizing the limitations in predicting intraday trading volume ratios due to their unavailability during trading hours and inherent variability, our approach focuses on stabilizing these ratios through log transformation~\cite{sumeth_tuvadaratragool_2023}, thereby yielding a more consistent log-normal distribution. This strategy not only aligns with the intricacies of VWAP tracking strategies but also addresses the challenges posed by the dynamic nature of trading volumes.

Delving deeper into the nuances of trading volume analysis, it becomes apparent that the simple statistical characteristics of volume alone are insufficient to capture the complexity of market behaviors~\cite{Graczyk_2016}. Our exploration reveals that trading volume curves are significantly shaped by a myriad of factors, including intraday seasonalities, nonstationarities~\cite{Graczyk_2016}, and absolute time information~\cite{Libman2019}. Furthermore, the liquidity levels of stocks introduce another layer of complexity, as evidenced by the disparate volume curves observed in stocks with varying liquidity~\cite{markov2019quintetvolumeprojection}. The consistent pattern of trading volume curves, despite these variations, emphasizes the importance of considering historical trading data, as evidenced in previous studies~\cite{Liu2016IntradayVP,markov2019quintetvolumeprojection}.

This understanding guided the development of our model, named IVE (Intraday Volume Estimator), with its input features tailored to maximize the efficiency of the Transformer Architecture. By integrating Time Encoding alongside the standard sinusoidal positional embedding, the model gains enhanced capabilities in processing absolute time information. The decision to set the context length at 390 one-minute steps, encompassing a full day's historical data, stems from the intention to provide a comprehensive snapshot of trading activities. This includes an array of volume-related features such as volume, accumulated volume, turnover rate, and trade amount, enabling a more nuanced understanding of the market dynamics.

The model’s ability to discern relevant predictive features from both the previous day and the current trading session is augmented by these curated features. Recognizing the variable scales of these features across stocks and dates, we adopted a normalization approach in line with established methodologies~\cite{Zhang_2019}, ensuring consistency and comparability across different datasets. Moreover, the inclusion of stock-specific categorical information, encoded and embedded within the model, allows for the nuanced capture of each stock's unique trading volume characteristics.

\subsection{Model Architecture and Training}
In this section, we delve into the model architecture and training methodology for predicting trading volume ratios using a transformer-based approach. Inspired by the Hugging Face blog~\cite{huggingface2022timeseries}, our model incorporates distribution heads into the transformer architecture to probabilistically predict trading volume ratios.

Input encoding methods for deep-learning approach for time series modeling usually encompass two main approaches: linear layers and convolutional layers. While 1D convolutional layers have been prevalent when modeling financial data with transformers~\cite{wallbridge2020transformerslimitorderbooks}, they are more effective for low-dimensional multivariate data. Given our focus on high-dimensional multivariate data, we opted for linear projection to encode input features~\cite{zerveas2020transformerbasedframeworkmultivariatetime}.

The transformed input features are subsequently processed by a transformer architecture. However, instead of a standard output layer, our model employs a distribution head that models the output as a Student's t-distribution. This choice is motivated by the observation from quantitative research~\cite{Graczyk_2016} that trading volume follows a fat-tailed stationary distribution in financial markets, making the t-distribution a more suitable choice compared to a normal distribution.

To prevent trivial solutions during training, we set our objective to predict not only one step ahead but up to three steps ahead. This approach addresses the tendency of time series transformers to impute missing values by predicting based on previous values~\cite{zerveas2020transformerbasedframeworkmultivariatetime}. Nevertheless, in practical applications, we utilize predictions up to one step ahead for trading decisions.

Our model was trained on data collected from January 25, 2023, to August 2023 from Refinitiv. This dataset includes the top 100 stocks by market capitalization in both the Korean and US markets as of September 2023. For each stock, we collected data on volume, accumulated volume, turnover rate, and the amount of trade. We used data from June 2023 as the validation set and data from August 2023 as the test set. June was chosen over July for the validation set due to its lower volatility, ensuring a more stable validation process.

Our model's training configuration involved a transformer architecture with both the encoder and decoder composed of four layers. The context length was set to one day, enabling the model to capture long-term dependencies effectively. For the output distribution, we employed a Student's t-distribution head, which is robust to outliers and heavy-tailed data.

We utilized the AdamW optimizer, which is known for its efficiency in handling sparse gradients and decoupling weight decay from the gradient update process. The learning rate was set to \(3 \times 10^{-4}\), a commonly used value that balances the convergence speed and stability.

Training was conducted on a high-performance computing setup with eight V100 GPUs, each with 32GB of memory, ensuring that the model could process large batches of data efficiently and leverage GPU parallelism to accelerate the training process.

To assess the effectiveness of our approach, we compare it with several baseline algorithms commonly used in previous research. These include bi-directional LSTM (Bi-LSTM), LSTM, and recurrent neural network (RNN). Previous studies often relied on hidden state models like Kalman filters, statistical models like ARMA and GARCH, and machine learning techniques like Gradient Boosting~\cite{satish2014, ma2021predictingdailytradingvolume, antulovfantulin2020temporalmixtureensemblemodels, Liu2016IntradayVP}. However, these methods were tailored to the statistical properties of trading volume curves, which cannot provide the context needed for predicting trading volume ratios.

LSTM-HR-AR, introduced in~\cite{Zhang_2019}, provided a notable precedent for using deep learning to predict trading volume ratios. However, the AR model in their architecture cannot be applied to our task of predicting trading volume ratios, leading us to select the LSTM-HR model for comparison. In the interest of a fair comparison, we constructed RNN and Bi-LSTM models similarly to the LSTM-HR model.

For our experiments, we collected data from the top 100 market capitalization stocks in both the Korean and US markets as of September 2023, conducting separate model training for each market. To evaluate model performance, we use Mean Absolute Error (MAE) and Root Mean Square Error (RMSE).

\begin{table}[h]
\centering
\begin{tabular}{lcc}
\toprule
\textbf{Model} & \textbf{RMSE} & \textbf{MAE} \\
\midrule
RNN-HR & 0.2540 & 0.1584 \\
LSTM-HR & 0.2295 & 0.1496 \\
BiLSTM-HR & 0.2055 & 0.1374 \\
IVE (our model) & \textbf{0.2028} & \textbf{0.1229} \\
\bottomrule
\end{tabular}
\caption{Experimental results for the Korean market (MAE in VWAP units).}
\end{table}

\begin{table}[h]
\centering
\begin{tabular}{lcc}
\toprule
\textbf{Model} & \textbf{RMSE} & \textbf{MAE} \\
\midrule
RNN-HR & \textbf{0.1665} & 0.1056 \\
LSTM-HR & 0.1690 & 0.1124 \\
BiLSTM-HR & 0.1725 & 0.1076 \\
IVE (our model) & 0.1678 & \textbf{0.0876} \\
\bottomrule
\end{tabular}
\caption{Experimental results for the U.S. market (MAE in VWAP units).}
\end{table}

Based on the provided experimental results for both the Korean and US markets, it is evident that the proposed model, IVE, demonstrates superior performance in predicting trading volume ratios. The effectiveness of IVE is particularly notable when comparing its Root Mean Square Error (RMSE) and Mean Absolute Error (MAE) metrics against other models such as RNN-HR, LSTM-HR, and BiLSTM-HR.

For the Korean market, IVE achieved the lowest RMSE of 0.2028 and MAE of 0.1229. This signifies a marked improvement over the BiLSTM-HR, which was the next best model with an RMSE of 0.2055 and MAE of 0.1374. The superior performance of IVE in this market can be attributed to its advanced architecture that efficiently captures the nuances of trading volume dynamics, a critical factor given the market's unique characteristics.

In the US market, the IVE also outperformed its counterparts, achieving an RMSE of 0.1678 and an MAE of 0.0876. Although the margin of improvement over other models like BiLSTM-HR (with an RMSE of 0.1725 and MAE of 0.1076) is narrower in the US market compared to the Korean market, it still underscores the effectiveness of IVE. The model's adeptness at handling the complexities of the US market, known for its high liquidity and volume, is a testament to the robustness of its design and the appropriateness of the chosen features and model architecture.

The consistency of IVE's performance across both markets suggests that its design is well-suited for diverse trading environments. This is an important finding, as it indicates that the model's approach to feature integration, time encoding, and the use of a Transformer-based architecture with a T-distribution head is universally effective. Furthermore, the decision to predict up to three steps ahead, as opposed to just one, seems to provide a more accurate representation of the dynamic nature of trading volumes, contributing to the model's enhanced performance.

\begin{figure}
  \centering
  \includegraphics[width=\columnwidth]{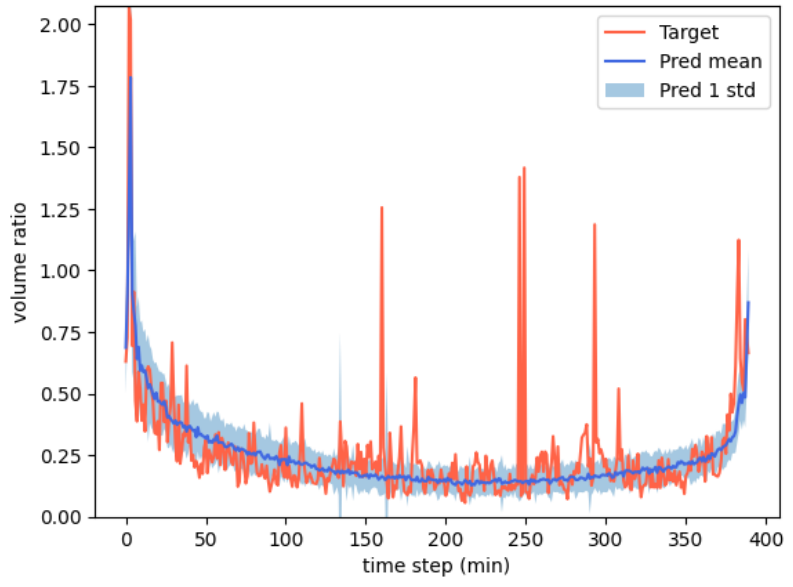}
  \caption{Example of Volume Ratio Prediction Result (AAPL).}
  \label{fig:single_column1}
\end{figure}

\begin{figure}
  \centering
  \includegraphics[width=\columnwidth]{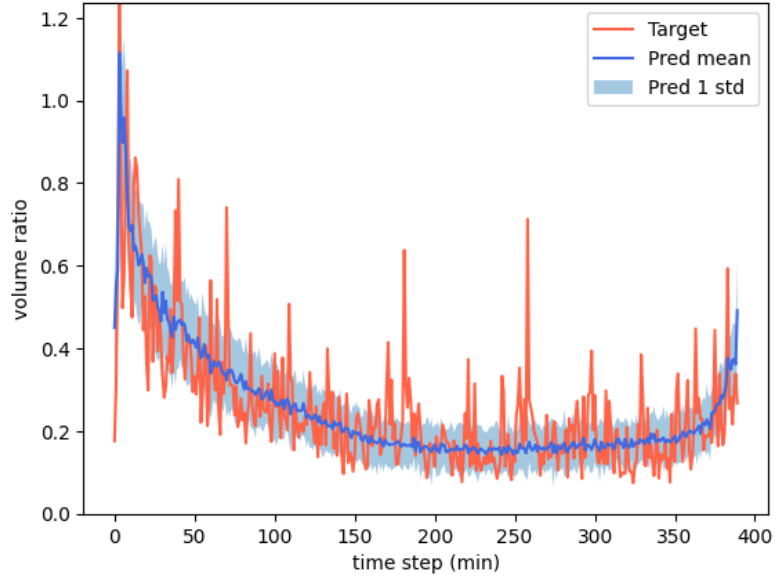}
  \caption{Example of Volume Ratio Prediction Result (TSLA).}
  \label{fig:single_column2}
\end{figure}

\section{Possibilities of Intraday Volume Spike Detection}
In the realm of financial market analysis, particularly in predicting volume ratios crucial for executing Volume-Weighted Average Price (VWAP) strategies, our study marks a significant departure from existing research which predominantly focused on point predictions. In this paper, we introduce a probabilistic model that not only predicts volume ratios but also incorporates statistical characteristics such as mean and standard deviation.

Underlying our research is the hypothesis that the standard deviation (std) of our model's output possesses predictive power regarding intraday volume ratio increases. To validate this premise, we employed the same test set used in previous performance evaluations, utilizing data from August 2023.

Our evaluation comprised two distinct phases. Initially, we investigated whether the predicted standard deviation holds a significant linear relationship with the volume ratio itself. Subsequently, we examined if the first difference of the predicted standard deviation has a meaningful linear relationship exclusively with positive values of the first-differenced volume ratio.

\begin{itemize}
    \item $p$-value: $3.56390 \times 10^{-123}$
    \item $R^2$: $0.07509$
    \item Coefficient: $0.06954$
\end{itemize}

These findings suggest a highly significant linear relationship between the predicted standard deviation and the volume ratio. The remarkably low p-value suggests that this relationship is not coincidental, while the R squared value implies that approximately 7.5\% of the total variability is explainable by the predicted standard deviation. The coefficient provides insight into the strength and direction of this linear relationship.

\begin{itemize}
    \item $p$-value: $3.92482 \times 10^{-52}$
    \item $R^2$: $0.08087$
    \item Coefficient: $0.08310$
\end{itemize}

Similarly, the results from the second experiment demonstrate a significant linear relationship between the standard deviation of predictions and the first difference of the volume ratio. The low p-value once again underscores the significance of these results, with the R squared value indicating that about 8\% of the variability is attributable to the predicted standard deviation.

The outcomes from these experiments underscore that the standard deviation of predictions has a meaningful correlation with market volatility, particularly in terms of volume. This correlation can signal increased uncertainty or volatility in the market, especially at points where the standard deviation is high. For investors, this could provide crucial signals indicating both risk and opportunity.

While the very low p-values affirm statistical significance, the relatively modest R squared values suggest that our model does not fully account for all variations in trading volume. This implies that other factors may influence market volume, suggesting the need for multivariate analysis or integration with other predictive indicators.

Integrating these predictions with other indicators could enhance the practical application of volume ratio prediction-based order execution algorithms. In our real-world trading tests, we conducted an analysis using historical data covering eight months to extract the median of volume ratio increases. We then developed a logic utilizing the model's standard deviation (std) specifically for cases exceeding this median. Notably, in these live trading tests, we observed that the model demonstrated predictive power where the increase in volume ratio was higher than the median. This observation offers insights into the practical utility of the model's uncertainty towards volume spikes in actual market scenarios. To account for sudden jumps in predicted values, we incorporated a constant factor of approximately 0.2, which was multiplied by the standard deviation and added to the predicted values. This constant was empirically adjusted based on observed performance.

\section{Real-World Trading and Outcomes}
In order to assess the practical utility of the Intraday Volume Estimator (IVE) model, we conducted live trading tests using an agent with a simple trading logic. The tests were carried out exclusively in the Korean market through the Korea Investment \& Securities Co. API, utilizing a seed capital of 350 million KRW.

\begin{algorithm}
\caption{Trading Strategy based on Predicted Volume Ratio}
\label{alg:trading_strategy}
\begin{algorithmic}[1]
\State \textbf{Input:} Predicted Volume Ratios $R$, Current Best Bid/Ask Prices $P$, Market Close Time $T_c$
\State \textbf{Output:} Executed Orders $O$
\State $O \gets \emptyset$
\For{each stock $s$ in selected stocks}
    \State Place order at the best bid/ask price $P_s$ based on predicted volume ratio $R_s$
\EndFor
\State $T_{cancel1} \gets T_c - 30$ minutes
\State $T_{cancel2} \gets T_c - 10$ minutes
\While{current time $t < T_{cancel1}$}
    \State Monitor order execution
\EndWhile
\For{each unfilled order $o$ in $O$}
    \State Cancel unfilled order $o$
    \State Place market order for remaining volume
\EndFor
\While{current time $t < T_{cancel2}$}
    \State Monitor order execution
\EndWhile
\For{each unfilled order $o$ placed after $T_{cancel1}$}
    \State Cancel unfilled order $o$
    \State Place final market order for remaining volume
\EndFor
\end{algorithmic}
\end{algorithm}

\subsection{Trading Strategy and Execution}
The trading strategy involved selecting five random stocks from the top 100 KOSPI stocks daily, excluding those with trading restrictions (such as company stocks or those traded the previous day). The trading logic was as follows:
\begin{itemize}
    \item Place orders at the best bid/ask based on the predicted trading volume ratios.
    \item Cancel all unfilled orders 30 minutes before market close and place market orders.
    \item Cancel all remaining unfilled orders 10 minutes before market close and place final market orders.
\end{itemize}

This strategy was executed from September 26, 2023, to November 23, 2023, resulting in a total of 163 orders across 3 to 5 stocks daily.

\subsection{Performance Metrics}
The performance of the executed trades was evaluated against the Market VWAP. The key metrics were:
\begin{itemize}
    \item \textbf{Average Execution Performance}: 4.82 basis points (bp) better than Market VWAP.
    \item \textbf{Standard Deviation of Execution Performance}: 34.59 bp.
    \item \textbf{Market VWAP Beat Ratio}: 59\%, indicating the strategy outperformed the Market VWAP more than half the time.
\end{itemize}

Furthermore, analyzing the top and bottom 20\% performance cases:
\begin{itemize}
    \item \textbf{Top 20\% Performance}: 15.55 bp better than Market VWAP.
    \item \textbf{Bottom 20\% Performance}: 10.08 bp worse than Market VWAP.
\end{itemize}

The results indicated that the majority of executions did not significantly deviate from the Market VWAP.

\subsection{Detailed Buy/Sell Performance}
A more granular look at the performance revealed differences between buy and sell orders:
\begin{table}[h]
\centering
\begin{tabular}{lcc}
\toprule
\textbf{Metric} & \textbf{Buy Orders} & \textbf{Sell Orders} \\
\midrule
Average Performance (bp) & 7.14 & 2.10 \\
Standard Deviation (bp) & 43.76 & 18.81 \\
\bottomrule
\end{tabular}
\caption{Performance Metrics of Buy and Sell Orders}
\end{table}

These differences suggest varying market conditions and volatility during the testing period.

The live trading tests showed that the IVE model can execute orders effectively, achieving performance comparable to VWAP strategies with a reasonable level of stability. The overall positive performance metrics confirm the model's practical utility in real-world trading scenarios.

\section{Conclusion}
In this study, we introduced the Intraday Volume Estimator (IVE), a Transformer-based model designed for enhanced probabilistic forecasting of intraday volume ratios in financial markets. The primary contributions of this work can be summarized as follows:

\textbf{Transformer-Based Volume Ratio Prediction Model:} We developed a model architecture leveraging the Transformer encoder-decoder framework, optimized for predicting minute-level intraday volume ratios. This model incorporates diverse features including statistical properties of volume, external volume-related features, absolute time information, and stock-specific characteristics. The introduction of a distribution head enables probabilistic forecasting, capturing both the mean and standard deviation of volume ratios.
    
\textbf{Probabilistic Forecasting Implementation:} Our model advances beyond traditional point predictions by providing probabilistic forecasts. This feature is crucial for anticipating intraday volume spikes and navigating market volatility. The relationship between predicted standard deviations and actual volume ratios has been empirically validated, showcasing the model's capability to signal increased market uncertainty and volatility.
    
\textbf{Practical Trading Application:} The IVE model's effectiveness was demonstrated through live trading tests in the Korean market, where it outperformed the VWAP benchmarks over a two-and-a-half-month period. The practical trading strategy, which relies on probabilistic volume forecasts, resulted in an average execution performance that was 4.82 basis points better than the Market VWAP, with a beat ratio of 59\%. This underscores the model's potential in real-world trading scenarios.

The results obtained from extensive experiments indicate that the IVE model offers improvements over existing methods in both the Korean and US markets. The model's robustness and adaptability to different market conditions highlight its practical utility and potential for widespread adoption in algorithmic trading.

Future research could explore several avenues to further enhance the performance and applicability of the IVE model. Integrating additional market indicators, and refining the probabilistic forecasting mechanism are promising directions. Moreover, the exploration of advanced optimization techniques for real-time trading strategy adjustments could yield even better execution outcomes.

In conclusion, our study presents an advancement in the field of intraday volume ratio prediction, integrating transformer architecture with practical trading applications.

\section*{Acknowledgments}
The authors would like to clarify that this research was conducted independently and is not affiliated with or influenced by the operations, interests, or policies of Shinhan Securities or Toss Bank. The views and conclusions expressed in this paper are solely those of the authors and do not reflect the official stance of their respective organizations.

\appendix

\section{Appendix}

\subsection{When is IVE More Useful?}
In this section, we investigate the specific market conditions under which the IVE demonstrates superior or inferior performance. The goal is to determine if certain features of the market can predict the performance of the algorithm on stock-date pairs. By defining market-explanatory features for each stock-date pair and performing multiple regression analysis, we aim to understand the impact of these features on the algorithm's performance.

\subsubsection{Data and Features}
We utilized execution performance data from September 26, 2023 to November 21, 2023. The features used in the analysis are as follows:
\begin{itemize}
    \item \textbf{open2close (x1)}: The ratio of the opening price to the closing price. A higher value indicates a significant increase during the trading day.
    \item \textbf{low2high (x2)}: The ratio of the lowest price to the highest price. A higher value suggests greater price volatility during the day.
    \item \textbf{min\_bar\_low2high\_mean (x3)}: The average ratio of the lowest to the highest price in each minute bar. Higher values indicate consistent price volatility within the minute bars.
    \item \textbf{min\_bar\_low2high\_std (x4)}: The standard deviation of the ratio of the lowest to the highest price in each minute bar. A higher standard deviation signifies frequent extreme price movements within the minute bars.
    \item \textbf{log\_turnover (x5)}: The logarithm of the turnover ratio (volume/number of outstanding shares). A higher value implies higher trading activity relative to the stock's size.
\end{itemize}

To compare the influence of these features on the algorithm's performance, we performed z-score normalization on the data before analysis.

\subsubsection{Multiple Regression Analysis Results}
The results of the multiple regression analysis are summarized in Table~\ref{tab:regression_results}.

\begin{table}[htbp]
    \centering
    \begin{tabular}{lrrrrr}
        \toprule
        Variable & Coefficient & Standard Error & t-Statistic & p-value \\
        \midrule
        const & 5.9946 & 2.691 & 2.194 & 0.030  \\
        x1 & 7.7286 & 4.316 & 1.791 & 0.075 \\
        x2 & -4.0735 & 8.201 & -0.497 & 0.620 \\
        x3 & -25.1314 & 6.663 & -3.772 & 0.000  \\
        x4 & 39.0623 & 8.119 & 4.811 & 0.000  \\
        x5 & 2.6769 & 4.317 & 0.620 & 0.536  \\
        \bottomrule
    \end{tabular}
    \caption{Multiple Regression Analysis Results}
    \label{tab:regression_results}
\end{table}

The regression model itself is statistically significant at the 0.05 level, as indicated by the F-statistic and associated p-value. The R-squared value, however, is relatively low (0.210), suggesting that other factors may also significantly influence performance and should be explored in future research.

\subsubsection{Interpretation of Results}
\textbf{Market Conditions Favorable for IVE}:
\begin{itemize}
    \item \textbf{min\_bar\_low2high\_std (x4)}: The algorithm performs better in conditions of high volatility within minute bars. This implies that IVE is effective in environments with frequent extreme price movements.
    \item \textbf{open2close (x1)}: The algorithm tends to perform well on days with a general upward trend, as indicated by a positive coefficient for the opening to closing price ratio.
\end{itemize}

\textbf{Market Conditions Challenging for IVE}:
\begin{itemize}
    \item \textbf{min\_bar\_low2high\_mean (x3)}: The algorithm underperforms in scenarios where there is consistent price volatility within minute bars. This indicates a potential weakness in handling steady, high volatility throughout the trading day.
\end{itemize}

Despite the statistically significant model, the relatively low R-squared value highlights the need for further investigation into additional factors that could explain variations in algorithm performance.

\subsection{Performance under Volatility Interruption}
To evaluate the performance of the IVE during instances of Volatility Interruption (VI), we conducted a series of experiments using a simulator. The experiments were based on data from the Korean market, spanning from January 1, 2022, to October 30, 2023. We extracted the stock-date pairs where a VI occurred, focusing on 500 stocks.

The criteria for identifying VI cases were as follows:
\begin{itemize}
    \item High price being 10\% or more above the opening price.
    \item Low price being 10\% or more below the opening price.
\end{itemize}
This resulted in a total of 4,125 VI cases.

The test results showed a significant decline in performance during VI periods:
\begin{itemize}
    \item \textbf{ASK Performance:} On average, 197.456 basis points (bp) worse than VWAP.
    \item Median performance relative to VWAP: 159.179 bp worse.
    \item Standard deviation of performance relative to VWAP: 296.220 bp.
    \item \textbf{BID Performance:} On average, 193.648 bp worse than VWAP.
    \item Median performance relative to VWAP: 151.074 bp worse.
    \item Standard deviation of performance relative to VWAP: 292.223 bp.
\end{itemize}

\begin{table}[h]
\centering
\begin{tabular}{lccc}
\toprule
\textbf{Performance} & \textbf{ASK} & \textbf{BID} \\
\midrule
Average (bp) & -197.456 & -193.648 \\
Median (bp)  & -159.179 & -151.074 \\
Standard Deviation (bp) & 296.220 & 292.223 \\
\bottomrule
\end{tabular}
\caption{Performance Metrics during Volatility Interruption}
\end{table}

These simulator-based results indicate that the proposed IVE model is not suitable during periods of VI. When deployed in real-world trading scenarios, it is crucial to incorporate additional logic to handle such extreme market conditions effectively.

\bibliographystyle{iclr2025_conference}
\bibliography{iclr2025_conference}

\end{document}

%% file: math_commands.tex
\usepackage{amsmath,amsfonts,bm}

\def\eqref#1{equation~\ref{#1}}
\def\1{\bm{1}}

\DeclareMathAlphabet{\mathsfit}{\encodingdefault}{\sfdefault}{m}{sl}
\SetMathAlphabet{\mathsfit}{bold}{\encodingdefault}{\sfdefault}{bx}{n}